\documentclass[preprint,12pt]{elsarticle}

\usepackage{amssymb}
\usepackage{amsmath}

\biboptions{sort&compress}
\usepackage{hyperref}
\usepackage[nolist]{acronym}
\usepackage{physics}

\journal{Computer Physics Communications}

\begin{document}

\begin{frontmatter}

\title{Grid-Free Evaluation of Phonon-Limited Electronic Relaxation Times and Transport Properties}

\author{Nenad Vukmirovi\'c}

\affiliation{organization={Institute of Physics Belgrade, University of Belgrade},
            addressline={Pregrevica 118},
            city={11080 Belgrade},
            country={Serbia}}

\begin{abstract}
Present calculations of electrical transport properties of materials require  evaluations of electron-phonon coupling constants on dense predefined grids of electron and phonon momenta and performing the sums over these momenta. In this work, we present the methodology for calculation of carrier relaxation times and electrical transport properties without the use of a predefined grid. The relaxation times are evaluated by integrating out the delta function that ensures energy conservation and performing an average over the angular components of phonon momentum. The charge carrier mobility is then evaluated as a sum over appropriately sampled electronic momenta. We illustrate our methodology by applying to the Fr{\"o}hlich model and to a real semiconducting material ZnTe. We find that rather accurate results can be obtained with a modest number of electron  and phonon momenta, on the order of one hundred each, regardless of the carrier effective mass.
\end{abstract}


\end{frontmatter}

\begin{acronym}
 \acro{DFT}{Density functional theory}
 \acro{DFPT}{density functional perturbation theory}
 \acro{PBE}{Perdew-Burke-Ernzerhof}
 \acro{GGA}{generalized gradient approximation}
 \acro{HH}{heavy hole}
 \acro{LH}{light hole}
\end{acronym}

\section{Introduction}
\label{sec:sec1}

Electron-phonon interactions play a crucial role in determining various physical properties and processes in materials \cite{rmp89-015003}, such as electrical and thermal conductivity \cite{rpp83-036501,prb94-201201,prb95-075206,prb104-085203}, relaxation of hot carriers \cite{nl17-5012}, conventional superconductivity \cite{prb82-184509,prb87-024505}, temperature dependence of the band gap \cite{jpcm9-2305,prb23-1495,prb27-4760,pccp25-29017}, and spin relaxation \cite{ncomm11-2780,prb98-125204}. \ac{DFT} \cite{PhysRev.136.B864, PhysRev.140.A1133,RevModPhys.64.1045} and \ac{DFPT} \cite{rmp73-515,prb55-10355} provide a framework for calculating electron-phonon coupling constants. However, calculating physical properties that depend on electron-phonon interactions remains challenging. This is because electron-phonon coupling constants often vary rapidly within the relevant parts of the Brillouin zone, and physical properties are typically described by sums over phonon and/or electron momenta. One therefore needs to know the electron-phonon coupling constants on a rather dense grid of electron and phonon momenta - much denser than the one that is needed to perform converged \ac{DFPT} calculations.

In last two decades, a methodology for obtaining the electron-phonon coupling constants on a dense grid from \ac{DFPT} calculation on a coarse grid was developed \cite{prb76-165108} and improved \cite{prl115-176401,prb92-054307}. It is based on representation of electronic wave functions in Wannier basis and phonon displacements in a local basis, followed by their transformation to Bloch function and plane wave basis. It is thus usually referred to as the Fourier-Wannier interpolation procedure.
Various physical properties were then computed using this methodology \cite{npjcm9-156,cpc181-2140,cpc264-107970,cpc295-108950,nl14-1113,prb91-235419,prb94-085204,prb95-075206,prb102-094314,prb106-235202,prap21-054017,acsel4-456,prl123-096602,prb100-085204,apl124-172103,prm3-034601,prb109-125203,prm8-L051001}. For example, to obtain carrier mobility in semiconducting materials, a coarse grid on the order of $\sim\!8^3$ is typically used, while dense grids ranging from $\sim\!45^3$ up to $\sim\!600^3$ were reported~\cite{prr3-043022,prb94-201201,prb95-075206,prb102-094308,prb102-125116}. Due to the need to compute the electron-phonon coupling constants and perform their summation on such a dense grid, the methodology is rather computationally demanding and it is a significant computational effort to apply it to a particular material. Hence, there are on-going research activities aimed at making such calculations less computationally demanding \cite{prb102-125116,ncomms12-2222,npjcm10-8,npjcm6-46}.

In this work, we demonstrate a grid-free procedure for calculation of relaxation times and electrical transport properties. The relaxation times involve the sum/integral over phonon momenta of a term that contains a delta function that ensures energy conservation in the phonon emission or absorption process. We perform this integral by integrating out the delta function and reducing it to a two-dimensional sum that is evaluated using a Monte Carlo procedure. Electrical transport properties such as carrier mobility or Seebeck coefficient then involve the sum over electronic momenta. We perform this summation by sampling the electronic momenta using a Monte Carlo procedure. The computation of electron-phonon coupling constants for given electron and phonon momenta (usually referred to as the $\vb{k}$ and $\vb{q}$ points, respectively) is still performed using the Fourier-Wannier interpolation procedure. However, it turns out that in our approach a rather modest number of $\vb{k}$ and $\vb{q}$ points (around 100 each) is sufficient to get converged results for relaxation times and carrier mobility in contrast to the need to use rather dense grids of $\vb{k}$ and $\vb{q}$ points in usual approaches with uniform or nonuniform grids. Such an advantage comes from the way our procedure is designed as it samples the most relevant $\vb{k}$ and $\vb{q}$ points in contrast to the straightforward use of predefined grids.

The paper is organized as follows. In Sec.~\ref{sec:sec2.1} we present the methodology for calculation of relaxation times and in Sec.~\ref{sec:sec2.2} we present the method for evaluation of transport properties such as the carrier mobility. In Sec.~\ref{sec:sec3.1}, we first illustrate the methodology by applying it to the Fr{\"o}hlich model where exact solutions are known and the results can be compared with them. Subsequently, in Sec.~\ref{sec:sec3.2} we apply the methodology to a real semiconducting material ZnTe and evaluate the electron and hole momentum relaxation times and mobility. We finalise the paper with the discussion and the conclusions in Sec.~\ref{sec:sec4}.

\section{Methods}
\label{sec:sec2}

\subsection{Calculation of relaxation times}
\label{sec:sec2.1}
Carrier momentum relaxation time $\tau_{n\vb{k}}$ of a carrier in band $n$ whose wave vector is $\vb{k}$ is given by the expression \cite{prb104-085203,Jacoboni,rpp83-036501,prb98-205202}
\begin{equation}
\begin{split}
\frac{1}{\tau_{n\vb{k}}}=&\frac{2\pi}{\hbar}\frac{1}{N_{\vb{q}}}\sum_{m,\vb{q}} \sum_{\lambda,\pm}
\abs{\gamma_{n\vb{k},m{\vb{k}\pm\vb{q}}}^{\lambda}}^2
\qty(n_{\lambda\vb{q}}+\frac{1}{2}\mp \frac{1}{2})
\times \\
&\delta\qty(\varepsilon_{n\vb{k}}\pm\hbar\omega_{\lambda\vb{q}}-\varepsilon_{m,\vb{k}\pm\vb{q}})
\qty(1-\frac{\vb{v}_{n\vb{k}}\cdot \vb{v}_{m,\vb{k}\pm\vb{q}}}{\abs{\vb{v}_{n\vb{k}}}\cdot \abs{\vb{v}_{m,\vb{k}\pm\vb{q}}}}).
\label{eq:mrt_rad66}
\end{split}
\end{equation}
In Eq.~\eqref{eq:mrt_rad66} the energy of electronic state in band $n$ whose wave vector is $\vb{k}$ is denoted as $\varepsilon_{n\vb{k}}$, the energy of a phonon in mode $\lambda$ with wave vector $\vb{q}$ is denoted as $\hbar\omega_{\lambda\vb{q}}$, the occupation number of that phonon state determined by the Bose Einstein distribution at a temperature $T$ is denoted as $n_{\lambda\vb{q}}=\qty(\exp\frac{\hbar\omega_{\lambda\vb{q}}}{k_BT}-1)^{-1}$. The delta function in the expression ensures energy conservation in the phonon absorption (the plus sign in the expression) or emission (the minus sign in the expression) process. $\gamma_{n\vb{k},m{\vb{k}\pm\vb{q}}}^{\lambda}$ are the electron-phonon coupling constants between the electronic states $n\vb{k}$ and $m{\vb{k}\pm\vb{q}}$ due to phonon state $\lambda\vb{q}$, while $\vb{v}_{n\vb{k}}=\frac{1}{\hbar}\pdv {\varepsilon_{n\vb{k}}}{\vb{k}}$ is the electronic band velocity of state $n\vb{k}$. The summation is performed over all possible $N_{\vb{q}}$ phonon momenta $\vb{q}$, phonon modes $\lambda$, bands of the final electronic state $m$, including the possibility of phonon absorption (the plus sign in the sum) and emission (the minus sign in the sum) processes.

The main challenge in evaluating the expression given by Eq.~\eqref{eq:mrt_rad66} comes from performing the summation over phonon momenta of the function that contains a delta function. Namely one has to perform the sum of the following type
\begin{equation}
S=\frac{2\pi}{N_{\vb{q}}} \sum_{\vb{q}} f\qty(\vb{q}) \delta\qty[g\qty(\vb{q})],
\label{eq:stype_rad66}
\end{equation}
which is in the same time equal to
\begin{equation}
S=\frac{V_u}{\qty(2\pi)^2} \int \mathrm{d}{\vb{q}} f\qty(\vb{q}) \delta\qty[g\qty(\vb{q})]
\label{eq:stype_rad66-2},
\end{equation}
where $V_u$ is the volume of the unit cell of the crystal.

Different approaches have been employed so far to evaluate the sums or integrals of the type given by Eqs.~\eqref{eq:stype_rad66} or \eqref{eq:stype_rad66-2}.
One possibility is to broaden the delta function by replacing it with a Gaussian or a Lorentzian function and perform the summation, which is an approach that has been so far widely used in the literature \cite{prb92-075405,prb94-201201,prb97-121201}. The challenge in such an approach is to properly choose the broadening parameter. If a large value of broadening is chosen, the summation can be straightforwardly performed but the result does not correspond to the result of the expression that involves a delta function. If a small value of broadening is chosen, one then needs a rather dense grid of $\vb{q}$ points to faithfully represent the broadened delta function and the summation might become impossible in practice due to extremely large number of $\vb{q}$ points in the summation.

Another possibility is to use the tetrahedron method~\cite{ssc9-1763,prb89-094515,prb102-094308,prb104-085203}. In this approach the $\vb{q}$ space is divided into tetrahedra and contributions to the integral from each tetrahedron are evaluated. To approximately evaluate the integral over the volume of the tetrahedron, the functions $f$ and $g$ are evaluated at the vertices of the tetrahedron and it is assumed that their values inside the tetrahedron can be obtained from linear interpolation of their values at the vertices. The integral over the volume of the tetrahedron can then be performed analytically. In this approach, the challenge is that the grid has to be dense enough so that the linear interpolation approximation becomes accurate. In practice, one has to increase the grid size until convergence is reached.

In this work, we exploit the fact that the delta function in the integral in Eq.~ \eqref{eq:stype_rad66-2} can be integrated out. This gives significant benefits as difficulties with proper representation of the delta function are avoided and the dimensionality of the integral is reduced. We obtain
\begin{equation}
    S=\frac{V_u}{\qty(2\pi)^2} \int \sin q_{\theta}\:\dd q_{\theta}\:\dd q_{\varphi}
    \int q_r^2 \dd q_r f\qty(q_r,q_{\theta},q_{\varphi}) \delta\qty[g\qty(q_r,q_{\theta},q_{\varphi})],
\end{equation}
where $q_r$, $q_{\theta}$, $q_{\varphi}$ are spherical coordinates of the vector $\vb{q}$.
By performing analytically the integration over $q_r$ we obtain
\begin{equation}
S=\frac{V_u}{\qty(2\pi)^2} \int \sin q_{\theta} \dd q_{\theta} \dd q_{\varphi} h\qty(q_{\theta},q_{\varphi}) \label{eq:sVu_rad66}
\end{equation}
with
\begin{equation}
\begin{split}
 h\qty(q_{\theta},q_{\varphi})=\sum_n q_r^{(n)}\qty(q_{\theta},q_{\varphi})^2   f\qty[q_r^{(n)}\qty(q_{\theta},q_{\varphi}),q_{\theta},q_{\varphi}]
  \frac{1}{\abs{\pdv{g}{q_r}}}_{\vb{q}=\qty[q_r^{(n)}\qty(q_{\theta},q_{\varphi}),q_\theta,q_\varphi]}
\label{eq:hfun-rad66}
\end{split}
\end{equation}
where $q_r^{(n)}\qty(q_{\theta},q_{\varphi})$ is the $n-$th zero of the function $g\qty(q_r,q_{\theta},q_{\varphi})$ regarded as a function of $q_r$ for given $q_{\theta}$ and $q_{\varphi}$.

In our implementation, we calculate the integral from Eq.~\eqref{eq:sVu_rad66} using a Monte Carlo procedure. Namely, we select $N_{\Omega}$ pairs $\qty(q_{\theta},q_{\varphi})$ as random points on a unit sphere and calculate the average
\begin{equation}
S=\frac{V_u}{\pi N_{\Omega}}\sum_{(q_{\theta},q_{\varphi})} h\qty(q_{\theta},q_{\varphi}).
\end{equation}
To evaluate the function $h$ using Eq.~\eqref{eq:hfun-rad66} one needs to find the zeros of the function $g\qty(q_r,q_{\theta},q_{\varphi})$ regarded as a function of $q_r$ (for fixed $q_{\theta}$ and $q_{\varphi}$). We accomplish this as follows. We search for the zeros in the interval $\qty(q_{\mathrm{min}},q_{\mathrm{max}})$, where $q_{\mathrm{min}}$ is some rather small value (typically $q_{\mathrm{min}}=10^{-4}\:\mathrm{bohr}^{-1}$) and $q_{\mathrm{max}}$ is the maximal value that still keeps the vector $\vb{q}=(q_r=q_{\mathrm{max}},q_{\theta},q_{\varphi})$ within the first Brillouin zone. We divide this interval into $N_1$ (typically $N_1=20$) subintervals in such a way that the subinterval boundary values form a geometric progression. For each of these subintervals, we check if the signs of the function $g$ at the boundaries are opposite. When this is the case, we search for the zeros of the function $g$ in the interval using the bisection method, where we typically apply $N_2=10$ iterations of the method.

\subsection{Calculation of charge carrier mobility}
\label{sec:sec2.2}
With carrier momentum relaxation times at hand, one can calculate various carrier transport properties, such as the charge carrier mobility, the Seebeck coefficient or the electronic thermal conductivity. Here we present our procedure for the case of charge carrier mobility. It is straightforward to apply the same procedure to calculate the other mentioned transport properties. The components of the charge carrier mobility tensor are given as \cite{prb104-085203,Jacoboni,rpp83-036501,prb98-205202}
\begin{equation}
\mu_{\alpha\beta}=-e_0\frac{\sum_{n\vb{k}} \pdv{f_{n\vb{k}}}{\varepsilon_{n\vb{k}}}\tau_{n\vb{k}}\:\qty(\vb{v}_{n\vb{k}})_{\alpha} \qty(\vb{v}_{n\vb{k}})_{\beta}}
{\sum_{n\vb{k}} f_{n\vb{k}}},
\end{equation}
where $f_{n\vb{k}}$ is the occupation of electronic state $n\vb{k}$ and the indices $\alpha$ and $\beta$ take the values $x$, $y$ or $z$, while $e_0$ is the elementary charge. In the typical scenario of low carrier concentration these occupations are at equilibrium given by the Maxwell-Boltzmann distribution $f_{n\vb{k}}\propto w_{n\vb{k}}=\exp{-\frac{\varepsilon_{n\vb{k}}-\varepsilon_m}{k_BT}}$. $\varepsilon_m$ is the reference energy that can be chosen arbitrarily, whereas it is most natural to choose it as the extremum of the most populated band (for example, the bottom of conduction band when we are interested in the mobility of electrons). The mobility is then given as
\begin{equation}
\mu_{\alpha\beta}=\frac{e_0}{k_BT}\frac{\sum_{n\vb{k}} w_{n\vb{k}} \tau_{n\vb{k}} \:\qty(\vb{v}_{n\vb{k}})_{\alpha} \qty(\vb{v}_{n\vb{k}})_{\beta}}
{\sum_{n\vb{k}} w_{n\vb{k}}}.\label{eq:mu2_rad66}
\end{equation}
The components of the mobility tensor $\mu_{\alpha\beta}$ can then be calculated as the average of the quantity $Q_{n\vb{k}}^{\alpha\beta}=\frac{e_0}{k_BT}\tau_{n\vb{k}} \:\qty(\vb{v}_{n\vb{k}})_{\alpha} \qty(\vb{v}_{n\vb{k}})_{\beta}$ with weights $w_{n\vb{k}}$. This average is calculated using a Monte Carlo procedure as follows. A random point $\vb{k}$ in the Brillouin zone is selected. A random band number $n$ is selected [from few (typically one to four) bands in the relevant energy range]. The state $n\vb{k}$ is accepted with a probability $w_{n\vb{k}}$. The procedure of selection of $n\vb{k}$ states is repeated and $\mu_{\alpha\beta}$ is eventually calculated as the average over all $N_k$ accepted $n\vb{k}$ states.

The described Monte Carlo procedure for evaluation of $\mu_{\alpha\beta}$ gives the result at a single given temperature $T$. In practice, one is usually interested in temperature dependence of the mobility. By repeating the procedure at a different temperature, different $n\vb{k}$ states would be sampled and hence one would need to obtain new sets of electron-phonon coupling constants to calculate $\tau_{n\vb{k}}$. This is not desirable due to the cost of calculation of electron-phonon coupling constants. It would be desirable to use the same electron-phonon coupling constants for calculation at all temperatures. This can be accomplished as follows.
Eq.~\eqref{eq:mu2_rad66} can be rewritten as
\begin{equation}
\mu_{\alpha\beta}\qty(T)=\frac{e_0}{k_BT}\frac{\sum_{n\vb{k}} w_{n\vb{k}}\qty(T_r)
\frac{w_{n\vb{k}}\qty(T)}{w_{n\vb{k}}\qty(T_r)}
\tau_{n\vb{k}}\qty(T) \:\qty(\vb{v}_{n\vb{k}})_{\alpha} \qty(\vb{v}_{n\vb{k}})_{\beta}}
{\sum_{n\vb{k}} w_{n\vb{k}}\qty(T_r)\frac{w_{n\vb{k}}\qty(T)}{w_{n\vb{k}}\qty(T_r)}},
\label{eq:mu_sam_rad66}
\end{equation}
where $T_r$ is some reference temperature that can be arbitrarily chosen. To evaluate $\mu_{\alpha\beta}\qty(T)$ we sample the states $n\vb{k}$ in accordance with the distribution $w_{n\vb{k}}\qty(T_r)$ and evaluate the ratio of the averages of quantities $R_{n\vb{k}}^{\alpha\beta}=\frac{e_0}{k_BT}\frac{w_{n\vb{k}}\qty(T)}{w_{n\vb{k}}\qty(T_r)}\tau_{n\vb{k}}\qty(T) \:\qty(\vb{v}_{n\vb{k}})_{\alpha} \qty(\vb{v}_{n\vb{k}})_{\beta}$ and
$S_{n\vb{k}}=\frac{w_{n\vb{k}}\qty(T)}{w_{n\vb{k}}\qty(T_r)}$
with weights $w_{n\vb{k}}\qty(T_r)$. The factors $\frac{w_{n\vb{k}}\qty(T)}{w_{n\vb{k}}\qty(T_r)}$ that appear in previous equations can be considered as correction factors that take into account the fact that the state sampling temperature $T_r$ is different than the system temperature $T$.

We also make use of the symmetry of the crystal as follows. We rewrite Eq.~\eqref{eq:mu2_rad66} as follows
\begin{equation}
\mu_{\alpha\beta}=\frac{e_0}{k_BT}\frac{1}{N_G}\frac{\sum_{n\vb{k}}\sum_{g\in G} w_{n,R(g)\vb{k}} \tau_{n,R(g)\vb{k}} \:\qty(\vb{v}_{n,R(g)\vb{k}})_{\alpha} \qty(\vb{v}_{n,R(g)\vb{k}})_{\beta}}
{\sum_{n\vb{k}} w_{n\vb{k}}}.\label{eq:mu5_rad66}
\end{equation}
where $N_G$ is the number of elements of the point symmetry group $G$ of the crystal, $g$ are its elements and $R(g)$ are their matrix representations. Making use of the equalities $w_{n,R(g)\vb{k}}=w_{n,\vb{k}}$, $\tau_{n,R(g)\vb{k}}=\tau_{n,\vb{k}}$ and
\begin{equation}
\qty(\vb{v}_{n,R(g)\vb{k}})_{\alpha}=\sum_{\beta}R(g)_{\alpha\beta}\qty(\vb{v}_{n,\vb{k}})_{\beta}
\end{equation}
we arrive at
\begin{equation}
\mu_{\alpha\beta}=\frac{e_0}{k_BT}\frac{1}{N_G}\frac{\sum_{n\vb{k}}\sum_{g\in G} w_{n\vb{k}} \tau_{n\vb{k}} \sum_{\gamma\delta} R(g)_{\alpha\gamma} R(g)_{\beta\delta}\:\qty(\vb{v}_{n\vb{k}})_{\gamma} \qty(\vb{v}_{n\vb{k}})_{\delta}}
{\sum_{n\vb{k}} w_{n\vb{k}}}.\label{eq:mu6_rad66}
\end{equation}
This symmetrization ensures that the final result for the mobility tensor $\mu_{\alpha\beta}$ is compatible with crystal symmetries.

\section{Results}
\label{sec:sec3}

\subsection{Fr\"{o}hlich model}
\label{sec:sec3.1}
In this section, we apply the methodology developed to a simplified model of the material. The assumptions of the model are as follows. There is a single electronic band with parabolic electronic dispersion $\varepsilon_{\vb{k}}=\frac{\hbar^2k^2}{2m_{\mathrm{eff}}}$, where $m_{\mathrm{eff}}$ is the effective mass of the electron. There is a single dispersionless phonon mode of energy $\hbar\omega_{\mathrm{LO}}$. Electron-phonon coupling is of Fr\"{o}hlich type
\begin{equation}
\abs{\gamma_{\vb{k},\vb{k}\pm\vb{q}}}^2=\frac{e_0^2\hbar\omega_{\mathrm{LO}}}{2V_u\epsilon_0}
\qty(\frac{1}{\epsilon_r^{\infty}}-\frac{1}{\epsilon_r^{\mathrm{st}}})\frac{1}{q^2}
\label{eq:frl_rad66}
\end{equation}
where $\epsilon_0$ is the vacuum permittivity, $\epsilon_r^{\infty}$ is the high-frequency relative dielectric constant and $\epsilon_r^{\mathrm{st}}$ is the static relative dielectric constant. For brevity, we will refer to this model as Fr\"{o}hlich model in what follows.

We choose to study this model to benchmark our methodology since for the Fr\"{o}hlich model analytic expressions for relaxation times exist and the expression for mobility reduces to one-dimensional integrals that can be straightforwardly  evaluated. Hence, accurate results for relaxation times and mobility in this model are known and it can serve as a reliable reference.  Fr\"{o}hlich model is also a rather good approximation for electrons at room temperature in polar semiconducting materials. In that case, relevant states are the states at the bottom of conduction band and the dominant electronic scattering mechanism is due to polar coupling to longitudinal optical phonons which is well described by coupling constants given in Eq.~\eqref{eq:frl_rad66}. Hence, by applying our methodology to the Fr\"{o}hlich model and comparing it to the reference results for the Fr\"{o}hlich model we can identify typical values of $N_k$ and $N_\Omega$ that are necessary to obtain accurate results.

Within the Fr\"{o}hlich model, the momentum relaxation time is given as
\begin{equation}
\tau_{\vb k}=\frac{1}{W_a+W_e},
\label{eq:tau1_rad66}
\end{equation}
where (see, for example, Supplemental Material in Ref.~\cite{prb104-085203})
\begin{equation}
W_a=\frac{1}{2\pi}\frac{n_{\mathrm{LO}}m_{\mathrm{eff}}C}{\hbar^3k}
\qty[1-\frac{\qty(k-\sqrt{C_1})^2}{2k\sqrt{C_1}}\ln\abs{\frac{k+\sqrt{C_1}}{k-\sqrt{C_1}}}]
\label{eq:tau2_rad66}
\end{equation}
is the term that comes from phonon absorption process and
\begin{equation}
W_e=\frac{1}{2\pi}\frac{\qty(n_{\mathrm{LO}}+1)m_{\mathrm{eff}}C}{\hbar^3k}\Theta\qty(C_2)
\qty[1-\frac{\qty(k-\sqrt{C_2})^2}{2k\sqrt{C_2}}\ln\abs{\frac{k+\sqrt{C_2}}{k-\sqrt{C_2}}}]
\label{eq:tau3_rad66}
\end{equation}
comes from phonon emission process. The symbol $n_{\mathrm{LO}}$ denotes the average number of phonons of energy $\hbar\omega_{\mathrm{LO}}$, $C=\frac{\hbar e_0^2\omega_{\mathrm{LO}}}{2\epsilon_0}
\qty(\frac{1}{\epsilon_r^{\infty}}-\frac{1}{\epsilon_r^{\mathrm{st}}})$, $C_1=k^2+\frac{2m_{\mathrm{eff}}\omega_{\mathrm{LO}}}{\hbar}$, $C_2=k^2-\frac{2m_{\mathrm{eff}}\omega_{\mathrm{LO}}}{\hbar}$ and $\Theta$ denotes the step function.
The mobility in the case of Fr\"{o}hlich model then reads
\begin{equation}
\mu=\frac{e_0}{k_BT}\frac{\hbar^2}{3m_{\mathrm{eff}}^2}
\frac{\int_0^{\infty} \dd k\: k^4 w_k \tau_k}{\int_0^{\infty} \dd k\: k^2 w_k}.
\end{equation}

Unless otherwise stated, we perform the tests for the following parameters of the Fr\"{o}hlich model: $m_{\mathrm{eff}}=0.117\: m_0$ (with $m_0$ being the free electron mass), $\hbar\omega_{\mathrm{LO}}=25.66\:\mathrm{meV}$, $\epsilon_r^{\mathrm{st}}=9.4$, $\epsilon_r^{\infty}=6.9$. These parameters correspond to ZnTe semiconducting material \cite{Adachi}, where the chosen value of effective mass corresponds to conduction band electron in that material.

\begin{figure}[t]
\centering
\includegraphics[width=0.5\textwidth]{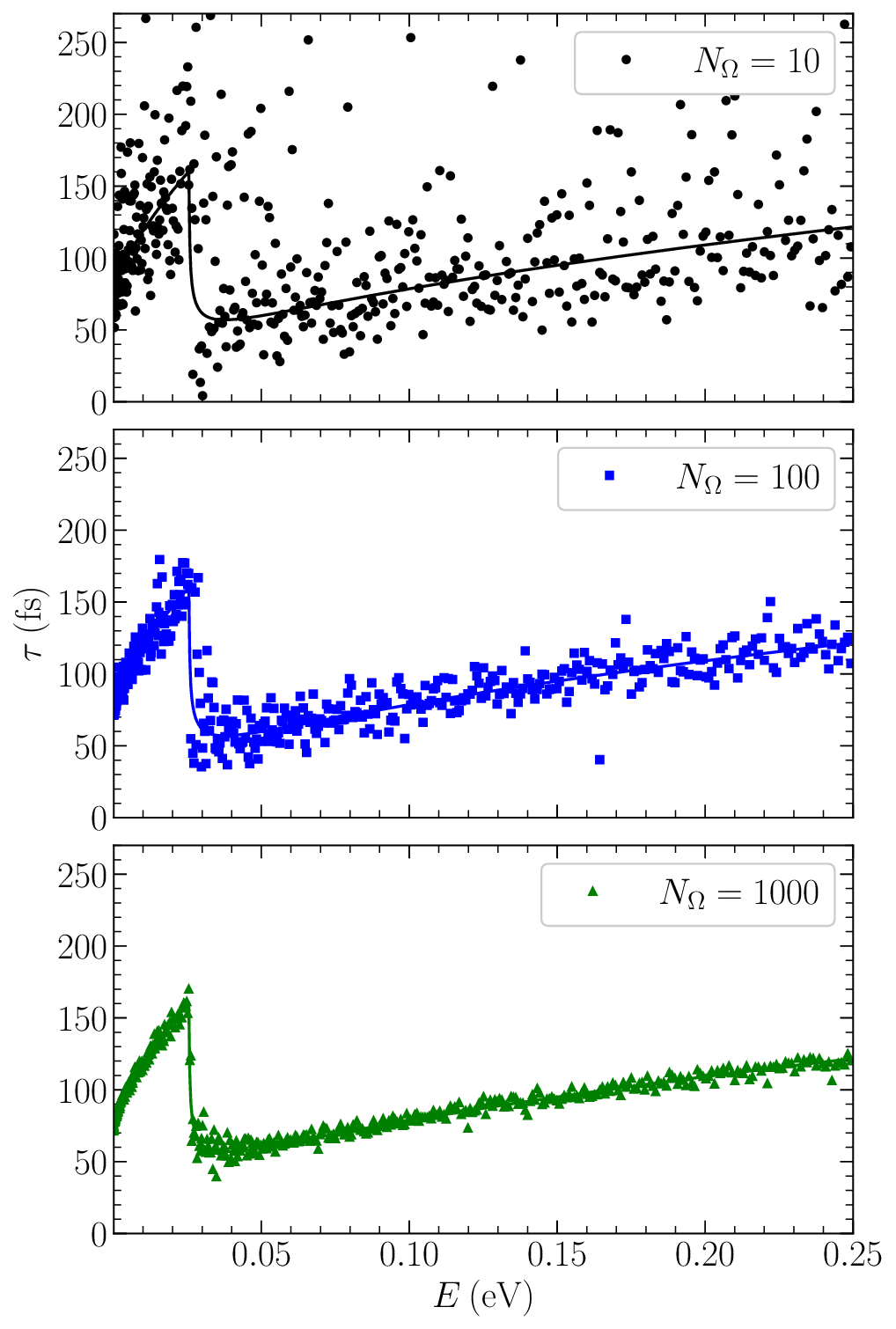}
\caption{Dependence of momentum relaxation time on electronic energy in the case of Fr\"{o}hlich model at $T=300\:\mathrm{K}$ for three different values of the number of points $N_{\Omega}$ used in the calculation. Full lines denote the reference analytical result.}\label{fig:a01}
\end{figure}

We start with calculations to determine necessary values of $N_{\Omega}$. In Fig.~\ref{fig:a01} we present the dependence of momentum relaxation time on electronic energy for three different values of $N_\Omega$ at $T=300\:\mathrm{K}$. The results are compared to analytical results that are calculated using Eqs.~\eqref{eq:tau1_rad66}-\eqref{eq:tau3_rad66} and presented in full lines in the figure.
The results obtained using $N_{\Omega}=10$ are rather inaccurate, while the results for $N_{\Omega}=100$ already follow nicely the analytical result and the results for $N_{\Omega}=1000$ almost fully agree with it.

\begin{figure}[t]
\centering
\includegraphics[width=0.5\textwidth]{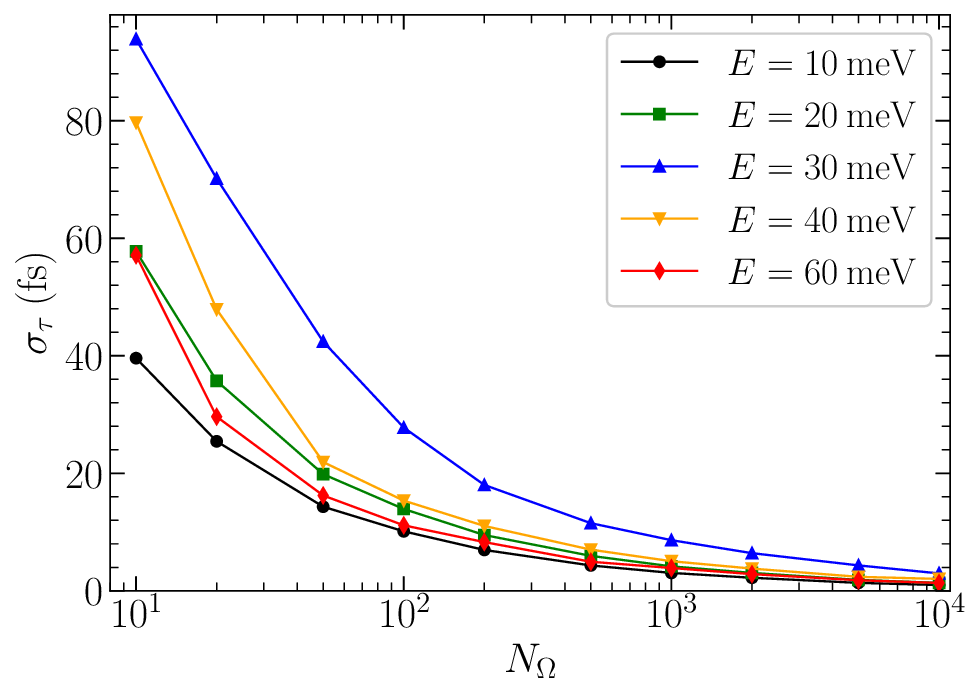}
\caption{The dependence of the standard deviation $\sigma_\tau$ of relaxation time on the number of points $N_\Omega$ used in its calculation. The results are presented for different energies of the electronic state $E$ and $T=300\:\mathrm{K}$.}\label{fig:a02}
\end{figure}

We describe the role of $N_{\Omega}$ more quantitatively by presenting the standard deviation of $\tau$ obtained from simulations with different random number seeds. The dependence of $\sigma_\tau$ on $N_{\Omega}$ for different values of electron energy $E$ and $T=300\:\mathrm{K}$ is presented in Fig.~\ref{fig:a02}. As expected $\sigma_\tau$ decays as $\propto\frac{1}{\sqrt{N_\Omega}}$. On the other hand, the dependence on $E$ for a fixed $N_\Omega$ is more complicated. It turns out that standard deviation is the largest at energies around $E=30\:\mathrm{meV}$, which is the energy just above the threshold energy for phonon emission (since $\hbar\omega_{\mathrm{LO}}=25.66\:\mathrm{meV}$). At these energies, the volume of the phase space of $\qty(q_\theta,q_\varphi)$ pairs for phonon emission is smallest and it is most difficult to sample this volume with a random choice of the $\qty(q_\theta,q_\varphi)$ pair.

\begin{figure}[t]
\centering
\includegraphics[width=0.5\textwidth]{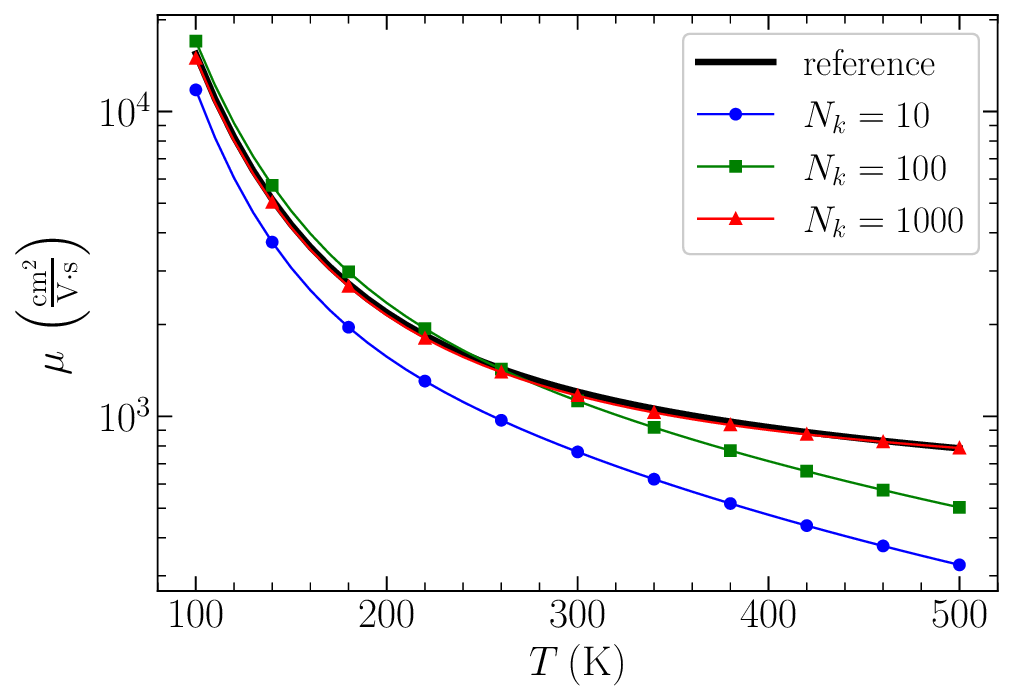}
\caption{Temperature dependence of the mobility within the Fr\"{o}hlich model for different numbers of sampling points $N_k$ and with reference temperature $T_r=300\:\mathrm{K}$. Full line denotes the accurate reference result.}\label{fig:a03}
\end{figure}

\begin{figure}[!h]
\centering
\includegraphics[width=0.5\textwidth]{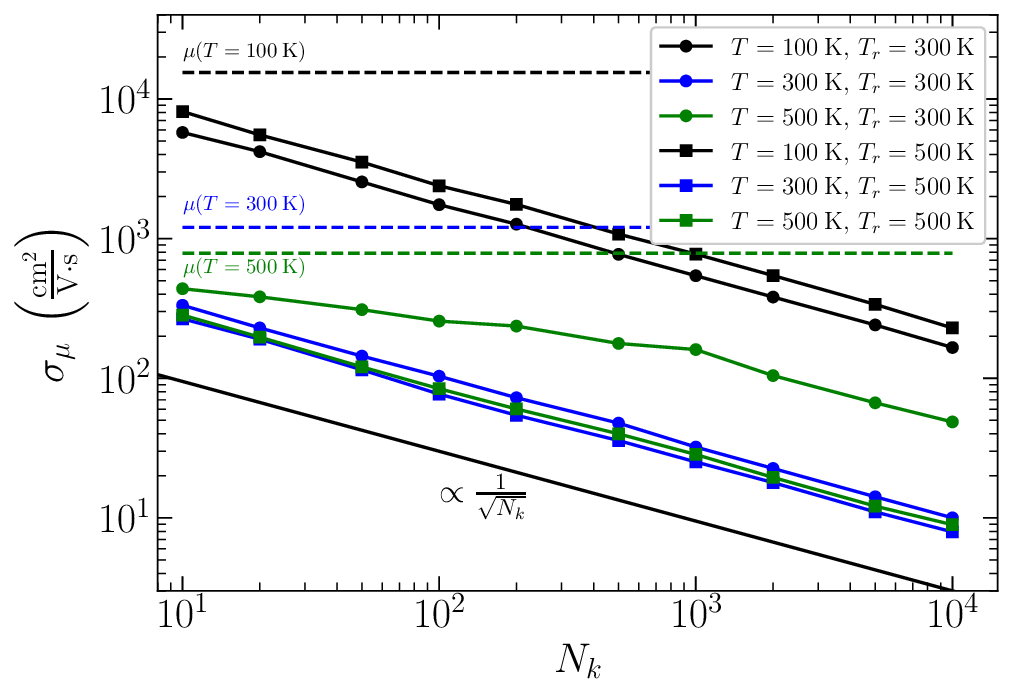}
\caption{Dependence of the standard deviation of mobility $\sigma_{\mu}$ on the number of $\vb{k}$ points $N_k$ used. The results are presented for different values of the temperature $T$ and reference temperature for sampling $\vb{k}$ points $T_r$. Horizontal dashed lines denote the values of the mobility at these temperatures.The line that describes the dependence $\propto\frac{1}{\sqrt{N_k}}$ is given as a guide to the eye.}\label{fig:a04}
\end{figure}

Next, we perform tests to understand the role of $N_k$ on uncertainty of the mobility results. For this purpose, we calculate relaxation times using analytical formulas from Eqs.~\eqref{eq:tau1_rad66}-\eqref{eq:tau3_rad66} and then perform the summation using Eq.~\eqref{eq:mu_sam_rad66}. That way, we isolate the effect of finite $N_k$ on the result. The result for the temperature dependence of mobility when the reference temperature of $T_r=300\:\mathrm{K}$ is used is presented in Fig.~\ref{fig:a03} for three different values of $N_k$. Results suggest that $N_k=1000$ is fully sufficient to get accurate results, while $N_k=10$ is clearly insufficient. For $N_k=100$ the results are quite accurate up to $T=300\:\mathrm{K}$.

To describe the role of $N_k$ more quantitatively, we repeat the calculation for different random number seeds and evaluate the standard deviation of the mobility $\sigma_{\mu}$. The results for different temperatures $T$ and for different value of the reference temperature $T_r$ are presented in Fig.~\ref{fig:a04}. An expected $\propto \frac{1}{\sqrt{N_k}}$ dependence of $\sigma_{\mu}$ is obtained for most $\qty(T,T_r)$ pairs [the exception for $\qty(T,T_r)=(500, 300)\:\mathrm{K}$ will be discussed in what follows] and we see that already at $N_k\sim 100$ the standard deviation is about an order of magnitude smaller than the mobility.

Next, we discuss the choice of the reference sampling temperature $T_r$. While it is most natural that this temperature is equal to the system temperature $T$, for reasons discussed in the previous section, it is beneficial to use a single $T_r$ for the whole interval of system temperatures of interest. While the final mobility result in the limit of large $N_k$ should not depend on $T_r$, the convergence towards that result with an increase of $N_k$ does depend on the choice of $T_r$. It is therefore of interest to understand the effect of the parameter $T_r$, that is, if it should be chosen from the middle or from the ends of the interval of interest. We see from Fig.~\ref{fig:a04} that as long as $T\le T_r$ there is only a modest effect of $T_r$ on $\sigma_{\mu}$ [compare the results for $\qty(T,T_r)=(100,300)\:\mathrm{K}$ to $\qty(T,T_r)=(100,500)\:\mathrm{K}$ or the results for $\qty(T,T_r)=(300,300)\:\mathrm{K}$ to $\qty(T,T_r)=(300,500)\:\mathrm{K}$]. On the other hand, when $T_r<T$ [the result for $\qty(T,T_r)=(500,300)\:\mathrm{K}$ in Fig.~\ref{fig:a04}] standard deviation becomes significantly larger [comparing to the case when $T_r=T$ is used, see the result for $\qty(T,T_r)=(500,500)\:\mathrm{K}$ in Fig.~\ref{fig:a04}] and the $\propto \frac{1}{\sqrt{N_k}}$ dependence is present only at high values of $N_k$ [on the order of $N_k\sim 1000$ for $\qty(T,T_r)=(500,300)\:\mathrm{K}$ in Fig.~\ref{fig:a04}]. Such a behavior suggests that the sampling temperature $T_r<T$ may not efficiently sample the region of higher electronic energies. This discussion therefore leads to the conclusion that for the calculation of mobility in an interval of temperatures, the temperature $T_r$ should be chosen as the highest temperature from that interval.

\begin{figure}[t]
\centering
\includegraphics[width=0.5\textwidth]{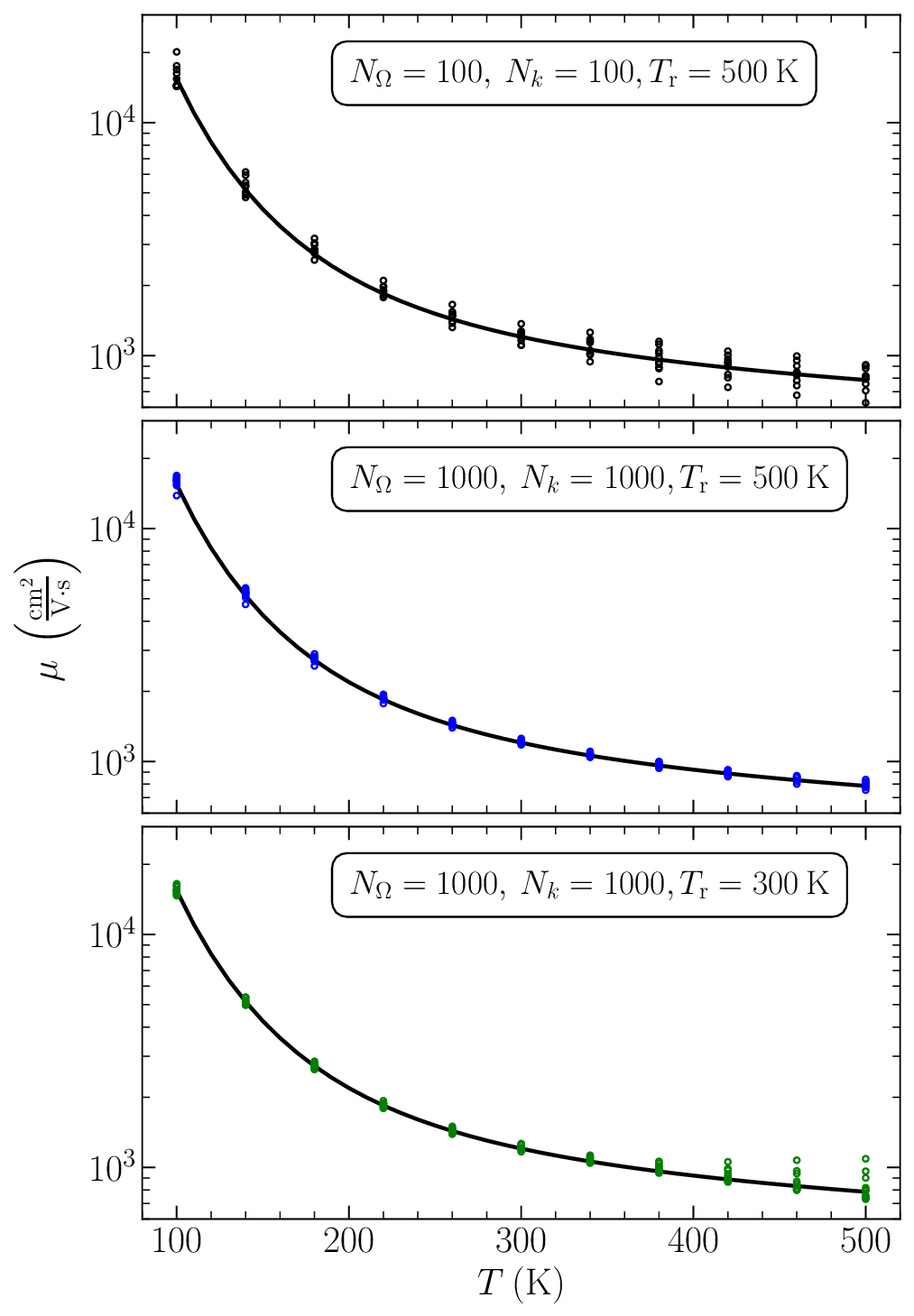}
\caption{Temperature dependence of mobility for the Fr{\"o}hlich model obtained by sampling $N_k$ $\vb{k}$-points using the sampling temperature $T_r$ and using $N_\Omega$ points for evaluation of relaxation times. The results are presented for ten calculations with different initial random number seeds. Full line denotes the accurate reference result.}\label{fig:a05_v2}
\end{figure}

In Fig.~\ref{fig:a05_v2} we present the final calculation result for the mobility obtained both by sampling $N_\Omega$ points for the evaluation of relaxation times and by sampling $N_k$ $\vb{k}$ points for the evaluation of the mobility. The results are presented for ten calculations with different initial random number seed.
It can be seen that the result for a rather modest number of points $N_\Omega=100$, $N_k=100$ already closely follows the exact result with reasonably small deviations for different calculations. The result for a larger number of points $N_\Omega=1000$, $N_k=1000$ (and $T_r=500\:\mathrm{K}$) matches the exact result almost perfectly (see middle part of Fig.~\ref{fig:a05_v2}). The result with the same number of points but using a smaller $T_r=300\:\mathrm{K}$ deviates from the exact result for temperatures that exceed $T_r$. This demonstrates again that the largest temperature from the interval of interest should be chosen for $T_r$.

We have thus established that, for the parameters of the Fr{\"o}hlich model used, rather modest values of $N_k$ and $N_\Omega$ are sufficient in our methodology to get accurate values of the mobility. We would like to establish if this conclusion can be extended regardless of the parameters of the model. We note first that the change in electron-phonon interaction only scales the relaxation times and the mobility and can therefore not change the convergence properties with respect to $N_k$ and $N_\Omega$. The two remaining parameters of the model are the effective mass $m_{\mathrm{eff}}$ and the phonon energy $\hbar\omega_{\mathrm{LO}}$. The phonon energy can be set as unit of energy and therefore we further discuss the role of $m_{\mathrm{eff}}$ on convergence properties.

In grid-based approaches to the problem, the effective mass crucially determines the size of the grid. This comes due to two effects: (i) A smaller effective mass leads to a larger band curvature and hence a denser grid of $\vb{k}$ points is needed to faithfully represent the electronic dispersion. For this reason, previous calculations have reported grids ranging from $45^3$ up to $600^3$ \cite{prr3-043022,prb94-201201,prb95-075206,prb102-094308} depending on the material considered; (ii) If an energy cut-off is introduced to consider only the electronic states up to some energy above/below the band minimum/maximum, a smaller effective mass leads to smaller volume of relevant $\vb{k}$-space. While the effect (i) tends to increase the needed number of $\vb{k}$ points for smaller effective mass, the effect (ii) tends to decrease it. Hence, it is generally not \textit{a priori} known which of these two effects will prevail. We have therefore performed the calculations for a different value of the effective mass $m_{\mathrm{eff}}=0.62\: m_0$ which corresponds to a spherically averaged mass of heavy holes in ZnTe \cite{Adachi}. The figures analogous to Figs.~\ref{fig:a03}-\ref{fig:a05_v2} are presented in Supplementary Material as Figs. S1-S3.
By comparing the two sets of figures, one finds that convergence properties are essentially the same for both values of the effective mass. Such a behavior can be understood as follows. Since $\vb{k}$ points are sampled based on their energies, for a parabolic dispersion $E=\frac{\hbar^2k^2}{2m_\mathrm{eff}}$, the $\vb{k}$ vectors and the distances $\Delta k$ between them will be proportional to $\propto\!\!\sqrt{m_{\mathrm{eff}}}$. Maximal typical intensities of $\vb{k}$ vectors that are sampled $k_m$ can be estimated from the relation $k_BT \propto \frac{\hbar^2k_m^2}{2m_{\mathrm{eff}}}$, which implies $k_m\propto \sqrt{m_{\mathrm{eff}}}$. The volume of the $\vb{k}-$space that is sampled is $V_k\propto k_m^3\propto m_{\mathrm{eff}}^{3/2}$. The number of $\vb{k}$ points that is sampled is then $N_k\propto V_k/\qty(\Delta k)^3\propto m_{\mathrm{eff}}^{3/2}/\qty(\sqrt{m_{\mathrm{eff}}})^{3}=1$, that is, it is not dependent on effective mass.

In fact, within our implementation of the procedure, when the calculations for effective masses $m_{\mathrm{eff}}^{(1)}$ and $m_{\mathrm{eff}}^{(2)}$ are performed with the same initial random number seed, the relation between $\vb{k}$ points that are sampled is $\vb{k}^{(1)}\qty[m_{\mathrm{eff}}^{(2)}]^{1/2}=\vb{k}^{(2)}\qty[m_{\mathrm{eff}}^{(1)}]^{1/2}$ and the mobilities obtained are exactly related as $\mu^{(1)} \qty[m_{\mathrm{eff}}^{(1)}]^{3/2}=\mu^{(2)} \qty[m_{\mathrm{eff}}^{(2)}]^{3/2}$. This can also be seen by comparing Figs.~\ref{fig:a03}-\ref{fig:a05_v2} to
Figs.~S1-S3
since the latter are simply scaled versions of the former with a factor equal to $\qty(m_{\mathrm{eff}}^{(2)}/m_{\mathrm{eff}}^{(1)})^{3/2}$.

\subsection{Real semiconductor material - ZnTe}
\label{sec:sec3.2}

In this section, we apply the methodology to a realistic semiconductor material - ZnTe.
As a first step in the calculation procedure, standard DFT calculation of the ZnTe material was performed. The material exhibits a zincblende crystal structure with room temperature experimental lattice constant of $6.0882\:\mathrm{\AA}$~\cite{Madelung}. This crystal structure was used as an input to our calculation. The \ac{PBE} \ac{GGA} \cite{prl77-3865} for the exchange-correlation potential was used.  The effects of core electrons were modeled using norm-conserving fully relativistic pseudopotentials \cite{prb88-085117,cpc226-39}. The effect of spin-orbit interaction was included. Plane waves with kinetic energy cut-off of $35\:\mathrm{Ha}$ were used to represent the electronic wave functions, while an $8\times 8\times 8$ grid of points in reciprocal space was used. Calculations were performed using the \textsc{ABINIT} code~\cite{cpc180-2582,cpc205-106,cpc248-107042,jcp152-124102}.

\begin{figure}[t]
\centering
\includegraphics[width=0.5\textwidth]{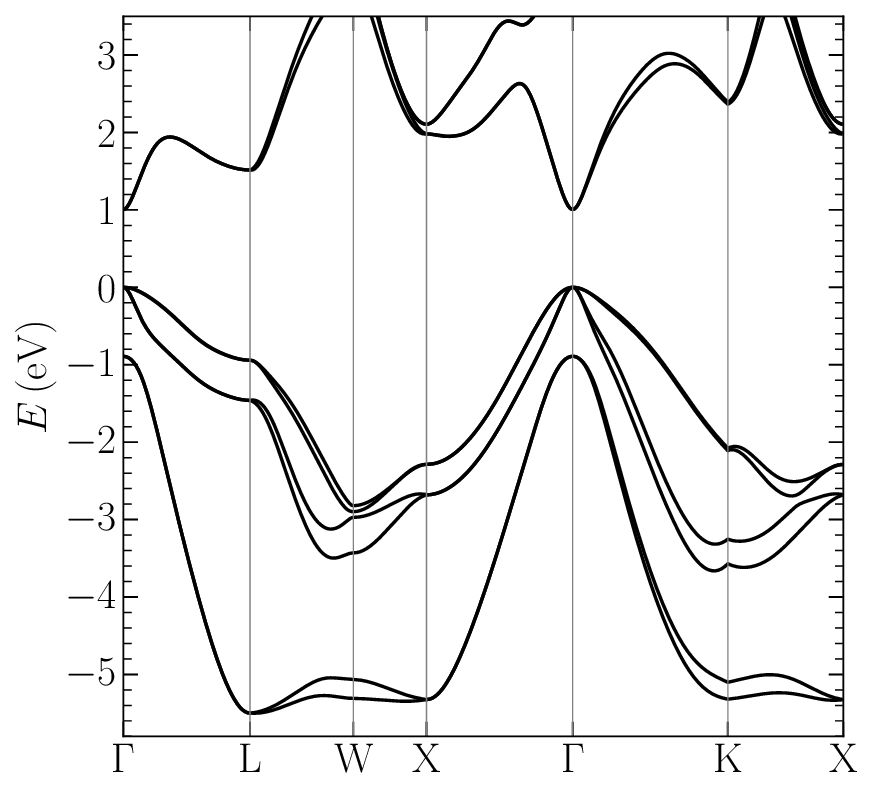}
\caption{Band structure of ZnTe along several relevant directions in the Brillouin zone. Valence band maximum energy is set as zero energy.}\label{fig:a06}
\end{figure}

We next describe the construction of Wannier functions. These were constructed starting from 36 Bloch wave functions from 18 highest valence bands and 18 lowest conduction bands. 18 Wannier functions were constructed. The procedure for construction of localized
Wannier functions for entangled energy bands \cite{prb65-035109} was used. In the band disentanglement procedure, the Bloch states that are within the energy window (so called frozen energy window \cite{cpc185-2309}) containing
the highest six valence bands and the two lowest conduction bands were left unchanged. \textsc{WANNIER 90} code \cite{cpc185-2309} used as a library within the \textsc{ABINIT} code was used in the construction of the unitary matrices that connect the Bloch and the Wannier functions. These matrices are then used to obtain the electronic energies and wave functions at arbitrary point in the Brillouin zone. The obtained band structure of ZnTe along several relevant directions in the Brillouin zone is presented in Fig.~\ref{fig:a06}.

The electron-phonon coupling constants and phonons (phonon energies and displacements of phonon modes) were calculated using \ac{DFPT}. The same kinetic energy cut-off and reciprocal space grid as for \ac{DFT} calculation was used. The \textsc{ABINIT} code was used for these calculations as well.

Interpolation of electron-phonon coupling constants and phonons to arbitrary wave vectors was performed using the Fourier-Wannier interpolation procedure, with proper treatment of long-ranged part of the electron-phonon coupling constants~\cite{prl115-176401,prb92-054307} and the dynamical matrices~\cite{rmp73-515,prb55-10355}. These calculations and subsequent calculations of carrier relaxation times and mobility were performed using our in-house code, which is an extension of the code used in Ref. \citep{prb104-085203}.

\begin{figure*}[t]
\centering
\includegraphics[width=0.65\textwidth]{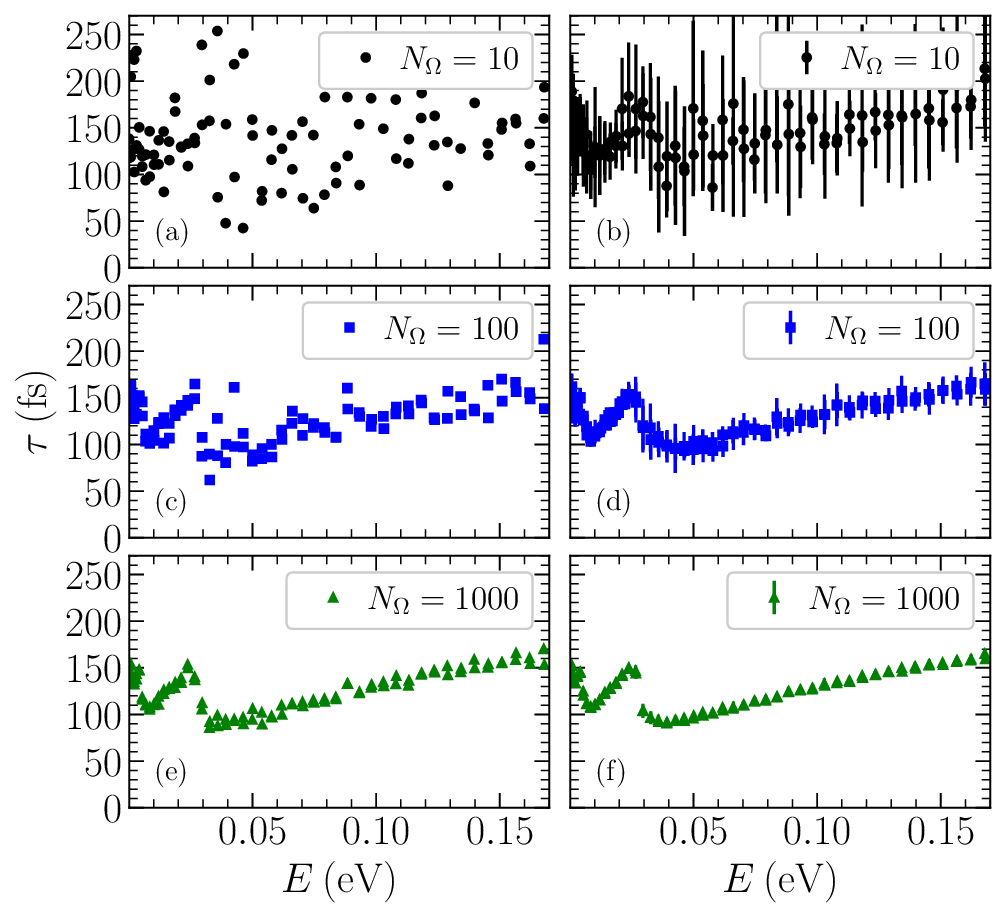}
\caption{Dependence of momentum relaxation time on the energy of an electron in conduction band of ZnTe for three different values of the number of points $N_{\Omega}$ used in the calculation at $T=300\:\mathrm{K}$. The electronic momentum is along the $\Gamma-L$ direction. The values shown in parts (a), (c) and (e) are the results for one realization.
The values shown in parts (b), (d) and (f) are the averages over ten different realizations, while the error bars are the standard deviation from these realizations. When error bars cannot be seen they are smaller than the symbol size.}\label{fig:a09}
\end{figure*}

As in the case of the Fr{\"o}hlich model, we start by investigating the role of $N_\Omega$ on the accuracy of the results. For this purpose we calculate the dependence of conduction band electron momentum relaxation time on energy of the carrier for different values of $N_\Omega$. In the calculation, we consider only the two lowest nearly degenerate conduction bands (see Fig.~\ref{fig:a06}) since these are the only ones relevant for relaxation of electrons at the bottom of the conduction band.
In Figs.~\ref{fig:a09}(a), (c) and (e) we present this dependence for a conduction band electron whose momentum is along the $\Gamma-L$ direction for three different values of $N_\Omega$ at $T=300\:\mathrm{K}$. For $N_\Omega=1000$ [Fig.~\ref{fig:a09}(e)] one sees a clear dependence of $\tau$ on $E$ with very little stochastic deviation from the trend implying that these results can be considered to be rather accurate. The results for $N_\Omega=100$ [Fig.~\ref{fig:a09}(c)] are overall similar to the ones for $N_\Omega=1000$ [Fig.~\ref{fig:a09}(e)], yet some deviations from the main dependence of $\tau$ on $E$ are noticeable. The results for $N_\Omega=10$ [Fig.~\ref{fig:a09}(a)] are clearly inaccurate since large deviations from the main trend are noticeable.

To quantify the comments from the previous paragraph, we repeated the $\tau\qty(E)$ calculation  ten times with different initial random number seed. We then calculated the average and the standard deviation of these ten realizations. The standard deviation then serves as a measure of accuracy of calculated $\tau$. The results of these calculations are presented in Figs.~\ref{fig:a09}(b), (d) and (f). The results for $N_\Omega=1000$ [Fig.~\ref{fig:a09}(f)] are highly accurate with typical standard deviation being only 3\% of the result. The results for $N_\Omega=100$ [Fig.~\ref{fig:a09}(d)] are also of rather good accuracy with typical standard deviations equal to around 10\% of the result. For $N_\Omega=10$ [Fig.~\ref{fig:a09}(b)] the results are not of sufficient accuracy, as typical standard deviation is around 30\% of the result in this case.

\begin{figure*}[t]
\centering
\includegraphics[width=0.65\textwidth]{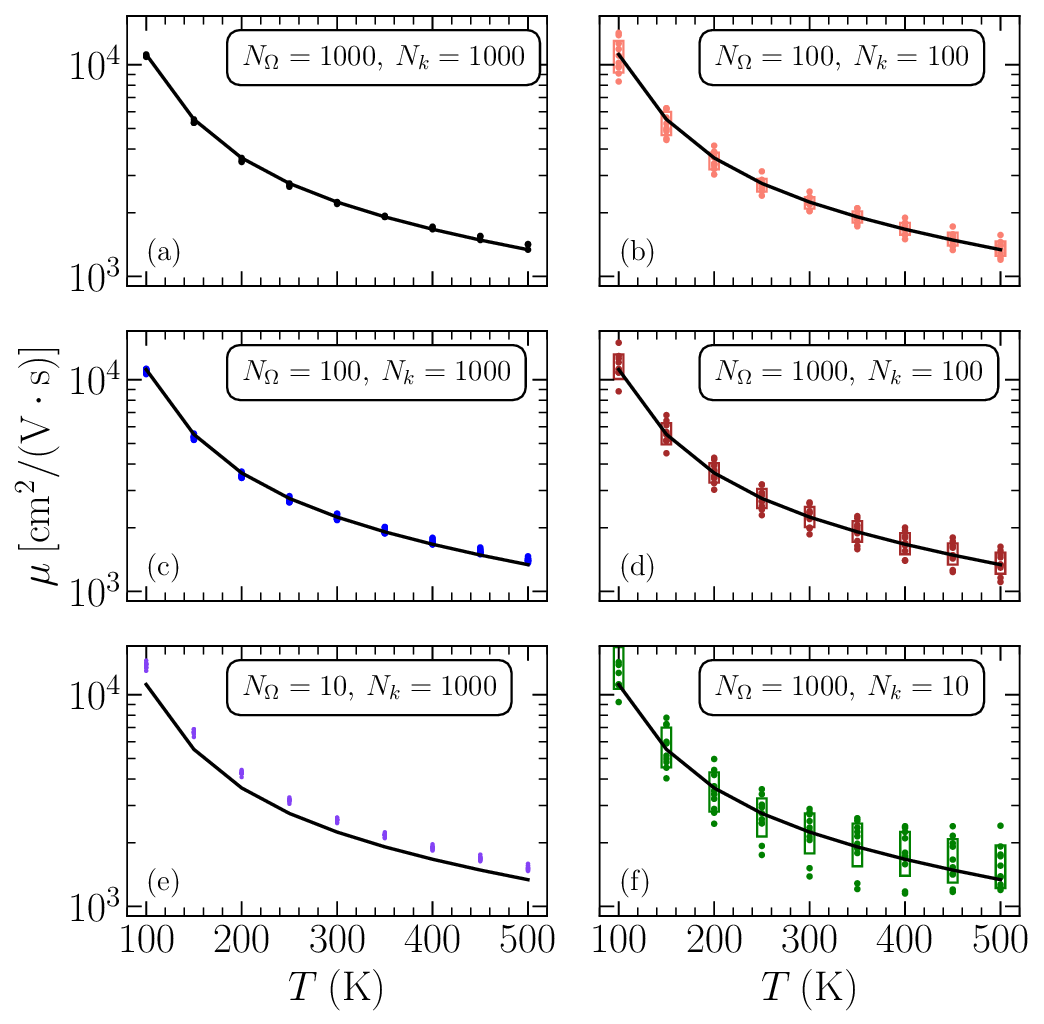}
\caption{Temperature dependence of electron mobility in ZnTe for different values of the number of $n\vb{k}$ pairs $N_k$ and the number $N_\Omega$. The dots present the results for ten (three when $N_k=1000$, $N_\Omega=1000$) different initial random number seeds. The rectangles denote the range $\qty(\mu-\sigma,\mu+\sigma)$ where $\mu$ is the average value and $\sigma$ is the standard deviation. Full line denotes the result for $N_k=1000$, $N_\Omega=1000$.}\label{fig:a11_CB}
\end{figure*}

We proceed next to analyze the influence of the number of $n\vb{k}$ pairs $N_k$ and the number $N_\Omega$ on the results for the mobility. We thus calculated the temperature dependence of the mobility for different
$\qty(N_k,N_\Omega)$ pairs. Based on previous experience with the Fr{\"o}hlich model we choose $T_r=500\:\mathrm{K}$ in all calculations. In Fig.~\ref{fig:a11_CB} we present the results for different values of $N_k$ and $N_\Omega$. For each $\qty(N_k,N_\Omega)$ pair we perform the calculation for ten different initial random number seeds. The results for these realizations are presented as dots in the figures.

We discuss first the influence of $N_\Omega$ on the results. We consider the results for  $N_k=1000$, $N_\Omega=1000$ as accurate reference results. Hence, we vary $N_\Omega$ while keeping $N_k=1000$ and analyze Figs.~\ref{fig:a11_CB}(c) and \ref{fig:a11_CB}(e). It can be seen from Fig.~\ref{fig:a11_CB}(e) that variations of the results for different realizations are rather small for $N_\Omega=10$, yet there is a systematic error due to small $N_\Omega$. This systematic error becomes rather small already at $N_\Omega=100$, as can be seen from Fig.~\ref{fig:a11_CB}(c).

Next, we consider the influence of $N_k$ on the results by keeping $N_\Omega=1000$. The results in Fig.~\ref{fig:a11_CB}(f) suggest that variations for $N_k=10$ are quite significant. When $N_k=100$ these variations fall to a relatively small level, with standard deviation being equal to 10-13\% of the results [Fig.~\ref{fig:a11_CB}(d)].

Overall, we thus find that modest values of $N_k$ and $N_\Omega$ of 100 are sufficient to get rather accurate results for the mobility. This can be seen from Fig.~\ref{fig:a11_CB}(b) where standard deviation of the result is below 20\% for $T<200\:\mathrm{K}$ and below 10\% for higher temperatures.
We consider the accuracy of 10-20\% as sufficient at present development of electronic structure methods. Namely, the effective masses calculated using standard functionals such as \ac{PBE} can introduce errors on the order of 30-50\% with respect to experimental values (see Ref. \cite{prb104-085203} for the case of II-VI semiconductors or \cite{prb106-045204} for a broader group of semiconductors) while these errors can be reduced to 10-20\% \cite{prb106-045204} when hybrid functionals (such as the HSE06 functional \cite{jcp118-8207,jcp125-224106}) or more advanced methods such as the many-body perturbation theory in GW approximation \cite{pr139-A796} are used. Hence, there is presently no need to  perform $\vb{k}$ and $\vb{q}$ summations with accuracy higher than the accuracy of electronic structure that is used as input. In our implementation, the calculation time for evaluation of mobility when  $N_k=100$ and $N_\Omega=100$ (excluding previous DFPT calculation and transformation from plane wave to local representation) is on the order of an hour using 64 CPU cores on a cluster with Intel Xeon  E5-2670 Processors.

Finally, we consider holes in the valence band of ZnTe. In Supplementary Material, we present figures analogous to Figs.~\ref{fig:a09}-\ref{fig:a11_CB} as Figs. S4-S5.
In these calculations, we consider the four highest bands - two nearly degenerate \ac{HH} and two nearly degenerate \ac{LH} bands (see Fig.~\ref{fig:a06}).
The dispersion of valence band holes is markedly different than the dispersion of conduction band electrons. Namely, the dispersion near the bottom of conduction band is isotropic with effective mass of $m_{\mathrm{eff}}=0.085\: m_0$ (extracted from band structure calculation using the \ac{PBE} functional). On the other hand, the top of the valence band accommodates anisotropic \ac{HH} and \ac{LH} bands (that are degenerate at the $\Gamma$ point) with \ac{PBE} effective masses of approximately $0.89\: m_0$ and $0.41\: m_0$ in [111] and [100] directions for \ac{HH}, and $0.088\: m_0$ and $0.098\: m_0$ in [111] and [100] directions for \ac{LH} (Fig.~\ref{fig:a06}). Hence, calculations in the valence band of ZnTe present a rather different test for our calculation methodology. Despite such a different dispersion, after analysis of
Figs.~S4-S5
we reach essentially the same conclusions regarding the influence of the parameters $N_k$ and $N_\Omega$ on the results. Namely, we find again that typical standard deviations of $\tau$ are 10\% for $N_\Omega=100$ and 3\% for $N_\Omega=1000$. When the mobility is concerned, we find again that modest values of $N_k=100$, $N_\Omega=100$ are sufficient to get rather accurate results for the hole mobility - its standard deviation is between 10\% and 15\% in this case.

\section{Discussion and Conclusions}
\label{sec:sec4}

Next, we discuss typical grid sizes and the number of $\vb{k}$ and $\vb{q}$ points that are required in other approaches for calculation of relaxation times and carrier transport properties and compare and contrast them to our approach. The need for a dense grid of $\vb{k}$ points comes from the fact that the carriers at energies of several $k_BT$ from the bottom of the conduction band (or the top of the valence band) are the ones that are relevant for transport properties. One then needs to faithfully represent the  electronic dispersion of carriers in this energy range, calling for a dense $\vb{k}$ point grid. To faithfully describe phonon-assisted transitions between the states in this energy range, a $\vb{q}$ point grid comparably dense to the $\vb{k}$ point grid is then needed.

As already mentioned, the use of grids ranging from $\sim\!45^3$ up to $\sim\!600^3$ were reported in the literature~\cite{prr3-043022,prb94-201201,prb95-075206,prb102-094308}.
Full grids of such sizes would typically imply millions of $\vb{k}$ and $\vb{q}$ points and summations over grids with such a large number of points would be impossible in practice.

In practice, one reduces the number of $\vb{k}$ and $\vb{q}$ points by considering only the ones that lead to states in the relevant energy window and the number of $\vb{k}$ points is reduced by exploiting the symmetry of the crystal. For example, in Ref.~\cite{prr3-043022}, for the case of cubic BN, the $250^3$ $\vb{k}$-point grid was reduced to 12,390 by selecting a $0.3\:\mathrm{eV}$ wide energy window and exploiting the symmetry, whereas 814,981 $\vb{q}$ points were necessary. In Ref.~\cite{prb97-121201}, for silicon, nonuniform grids with 85,000 inequivalent $\vb{k}$ points and 200,000 inequivalent $\vb{q}$ points were used.

In the studies of Refs.~\cite{prr3-043022,prb97-121201} the summation was performed by replacing the delta function with its broadened version. A more favorable reduction in the number of $\vb{q}$ points can be accomplished when the tetrahedron method is used to perform the summation. In this case, one can reduce the summation only to tetrahedra intersecting one of the possible isosurfaces $g\qty(\vb{q})=0$ \cite{prb102-094308}. By combining this reduction with previous ideas (energy windows and symmetry), converged results were obtained in Ref.~\cite{prb102-094308} with relatively small number of $\vb{k}$ and $\vb{q}$ points. In case of electrons in silicon, it was reported that $45^3$ $\vb{k}$ points grid with only 29 irreducible $\vb{k}$ points, combined with $90^3$ $\vb{q}$ point grid (with less than 2\% of it selected, implying around 15,000 $\vb{q}$ points), was sufficient \cite{prb102-094308}. Somewhat larger grid was needed for electrons in GaP, while the calculation appeared to be most challenging for GaAs \cite{prb102-094308} that has a low effective mass, implying the need for a rather dense grid. In our previous study \cite{prb104-085203} where we used the tetrahedron method, we obtained converged results for electrons in ZnTe with the $120^3$ grid, the energy window of $0.25\:\mathrm{eV}$, which lead to around 100 inequivalent $\vb{k}$ points and around 2000 to 5000 $\vb{q}$ points for each $\vb{k}$ point. Previous studies using the tetrahedron method thus suggest that the required grid size is heavily dependent on the band curvature (effective mass). We also find that it is rather challenging to apply the tetrahedron method in the case of holes when both \ac{HH} and \ac{LH} bands with rather different effective masses exist. In that case, the \ac{LH} band with small effective mass calls for the dense grid, while the \ac{HH} band with large effective mass leads to larger volume of the relevant part of the Brillouin zone. Combining these two requirements leads to the need for a rather large number of $\vb{k}$ and $\vb{q}$ points. Interestingly, we are not aware of a study where the tetrahedron method was used in the case of holes when both \ac{HH} and \ac{LH} bands exist.

In the approach presented in this work, we find that the number of required $\vb{k}$ and $\vb{q}$ points is much smaller or in worst case comparable to what was required in previous approaches. Our selection of $\vb{k}$ points based on the weights $w_{n\vb{k}}$ selects the relevant states at the bottom of conduction band (top of valence band) and produces a similar effect as the selection based on an energy window. The difference is that there is no sharp cut-off energy in our approach and therefore there is no need for a convergence study with respect to that cut-off parameter. The main advantage of our approach comes from the way the summation/integration over $\vb{q}$ points is performed. Energy conservation in the delta function effectively determines $q_r$ - the modulus of $\vb{q}$ and one is left with the integral over angular components $q_\theta$ and $q_\varphi$. This makes the number $N_\Omega$ of necessary $\qty(q_\theta,q_\varphi)$ pairs significantly smaller than the number of $\vb{q}$ points in other approaches. Another advantage of our approach is that convergence properties appear to be insensitive to band curvature and hence the approach can be applied with equal success to different materials, in contrast to grid-based methods where the number of required $\vb{k}$ and/or $\vb{q}$ points is heavily dependent on band curvature near the band extremum.

In conclusion, we presented a procedure for calculation of relaxation times and transport properties without the use of predefined $\vb{k}$ and $\vb{q}$ points grids. Numerical tests show that
converged results for carrier mobility with 10-20\% accuracy can be obtained with only 100 $\vb{k}$ points and 100 $\vb{q}$ points, regardless of band curvature at the extremal point. The procedure described should enable significantly more efficient \textit{ab-initio} calculations of electrical transport properties of semiconducting materials.

\section*{Data Statement}
The source codes with input and output files and the scripts used to prepare the figures are openly available at \url{http://doi.org/10.5281/zenodo.13772234}.

\section*{Acknowledgements}
This research was supported by the Science Fund of the Republic of Serbia, Grant No. 5468, Polaron Mobility in Model Systems and Real Materials--PolMoReMa.
The author acknowledges funding provided by the Institute of Physics Belgrade, through the grant from the Ministry of Science, Technological Development, and Innovation of the Republic of Serbia.
Numerical computations were performed on the PARADOX-IV supercomputing facility at the Scientific Computing Laboratory, National Center of Excellence for the Study of Complex Systems, Institute of Physics Belgrade.

\bibliographystyle{elsarticle-num}


\clearpage
\pagebreak
\newpage

\setcounter{equation}{0}
\setcounter{figure}{0}
\setcounter{table}{0}
\setcounter{page}{1}
\setcounter{section}{0}
\makeatletter
\renewcommand{\theequation}{S\arabic{equation}}
\renewcommand{\thefigure}{S\arabic{figure}}
\renewcommand{\bibnumfmt}[1]{[S#1]}
\renewcommand{\citenumfont}[1]{S#1}
\renewcommand{\thetable}{S\arabic{table}}
\renewcommand{\thepage}{S\arabic{page}}
\renewcommand{\thesection}{S\arabic{section}}

\begin{onecolumn}
\begin{center}
  \textbf{\large Supplementary Material: Grid-Free Evaluation of Phonon-Limited Electronic Relaxation Times and Transport Properties}\\[.2cm]
   Nenad Vukmirovi\'c,$^{1}$ \\[.1cm]
  {\itshape ${}^1$Institute of Physics Belgrade,
University of Belgrade, Pregrevica 118, 11080 Belgrade, Serbia}
  \\[1cm]
\end{center}


\section{Results for the Fr{\"o}hlich model with $m_{\mathrm{eff}}=0.62\: m_0$}

\begin{figure}[h]
\centering
\includegraphics[width=0.5\textwidth]{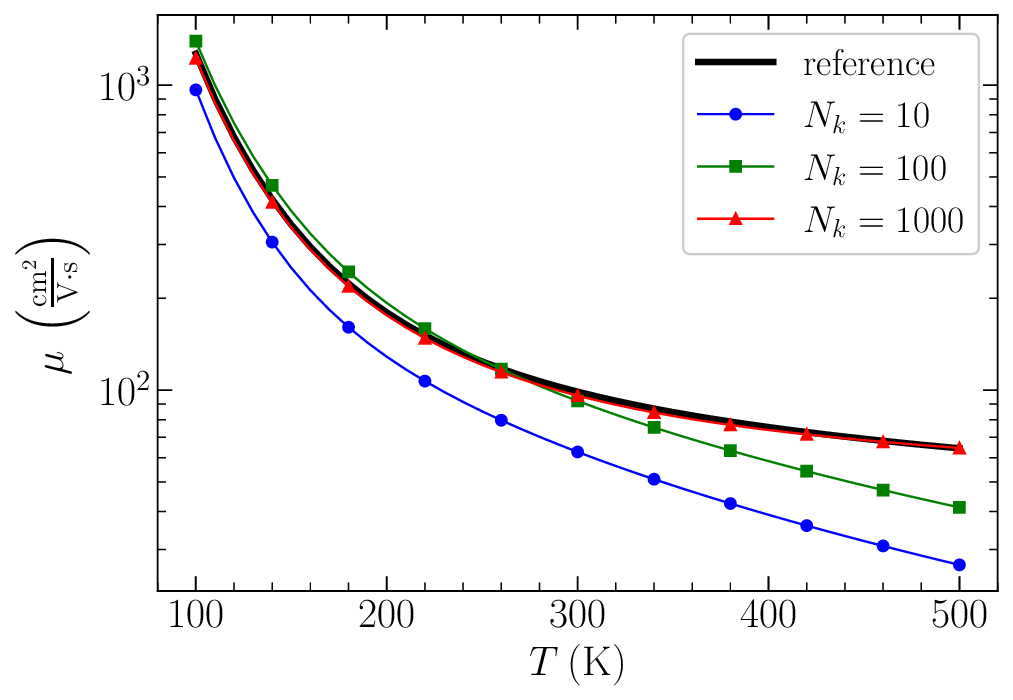}
\caption{Temperature dependence of the mobility within the Fr\"{o}hlich model with $m_{\mathrm{eff}}=0.62\: m_0$ for different numbers of sampling points $N_k$ and with reference temperature $T_r=300\:\mathrm{K}$. Full line denotes the accurate reference result.}\label{fig:a03_meff062}
\end{figure}

\begin{figure}[!h]
\centering
\includegraphics[width=0.5\textwidth]{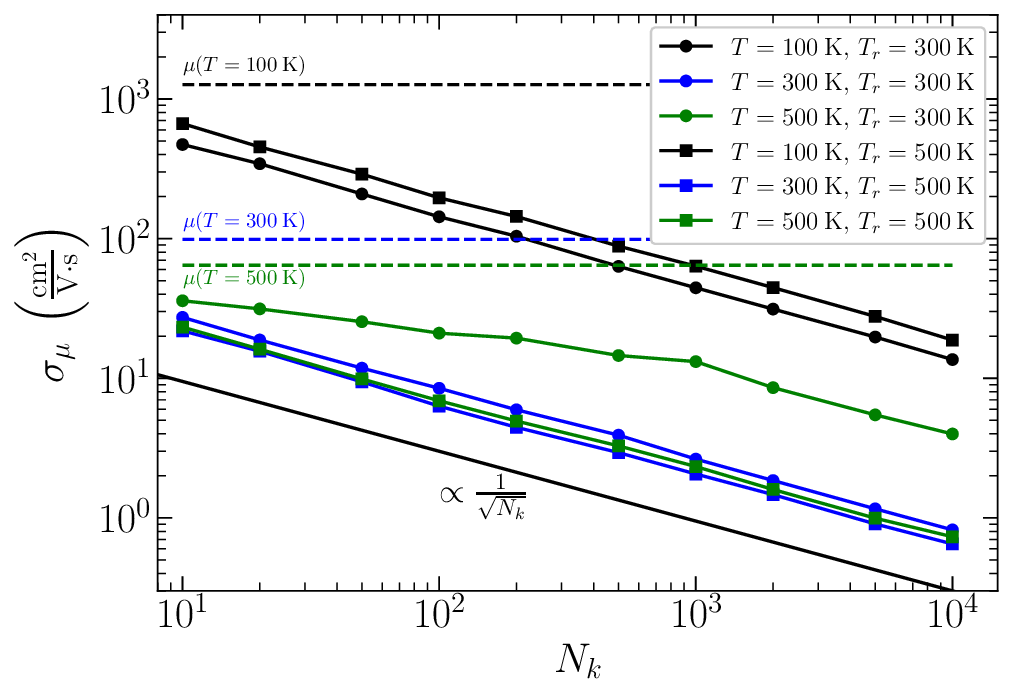}
\caption{Dependence of the standard deviation of mobility $\sigma_{\mu}$ on the number of $\vb{k}$ points $N_k$ used for the Fr{\"o}hlich model with $m_{\mathrm{eff}}=0.62\: m_0$. The results are presented for different values of the temperature $T$ and reference temperature for sampling $\vb{k}$ points $T_r$. Horizontal dashed lines denote the values of the mobility at these temperatures. The line that describes the dependence $\propto\frac{1}{\sqrt{N_k}}$ is given as a guide to the eye.}\label{fig:a04_meff062}
\end{figure}

\begin{figure}[t]
\centering
\includegraphics[width=0.5\textwidth]{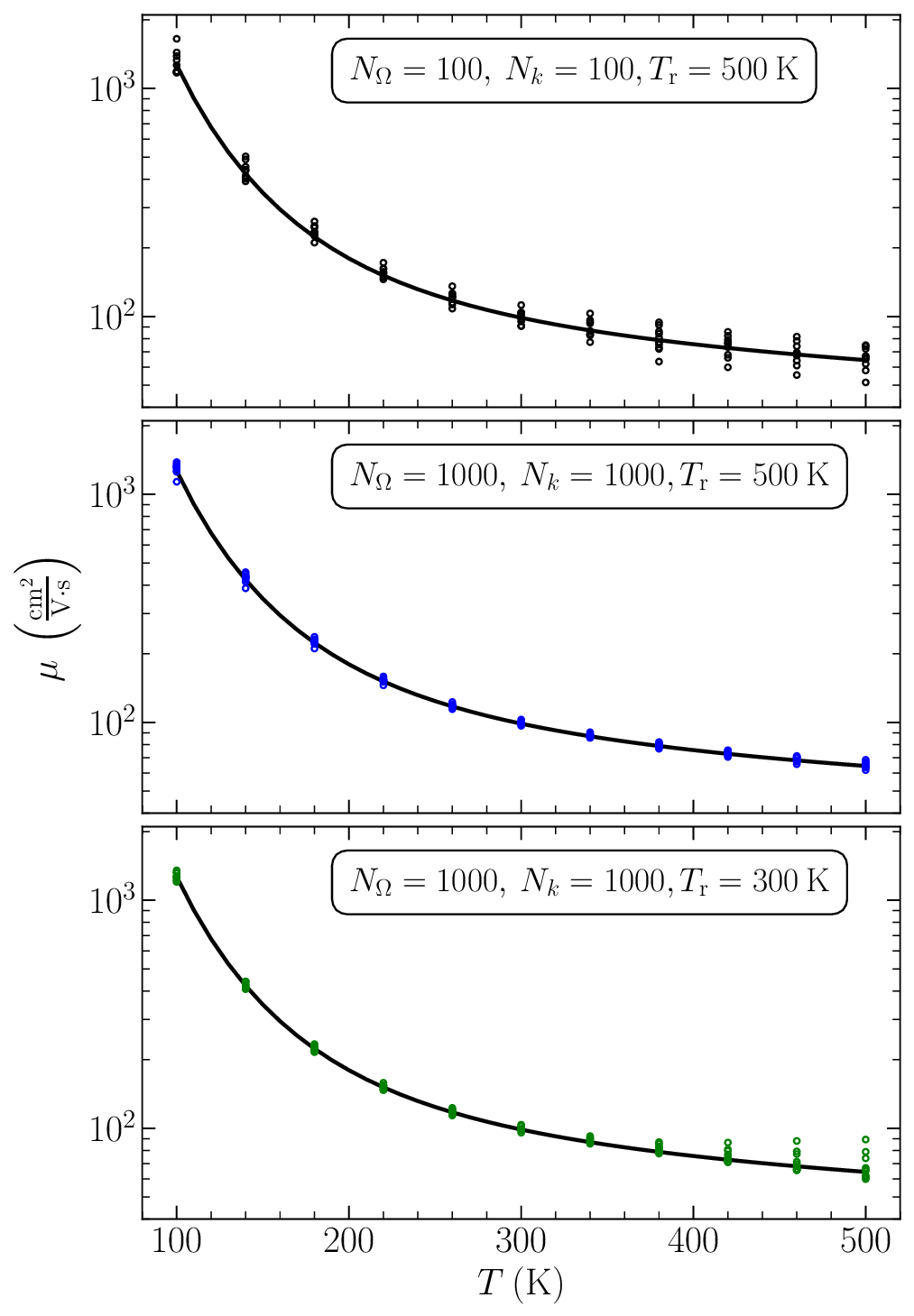}
\caption{Temperature dependence of mobility for the Fr{\"o}hlich model with $m_{\mathrm{eff}}=0.62\: m_0$ obtained by sampling $N_k$ $\vb{k}$-points using the sampling temperature $T_r$ and using $N_\Omega$ points for evaluation of relaxation times. The results are presented for ten calculations with different initial random number seeds. Full line denotes the accurate reference result.}\label{fig:a05_v2_meff062}
\end{figure}

\clearpage

\section{Results for holes in the valence band of ZnTe}

\begin{figure*}[htbp]
\centering
\includegraphics[width=0.65\textwidth]{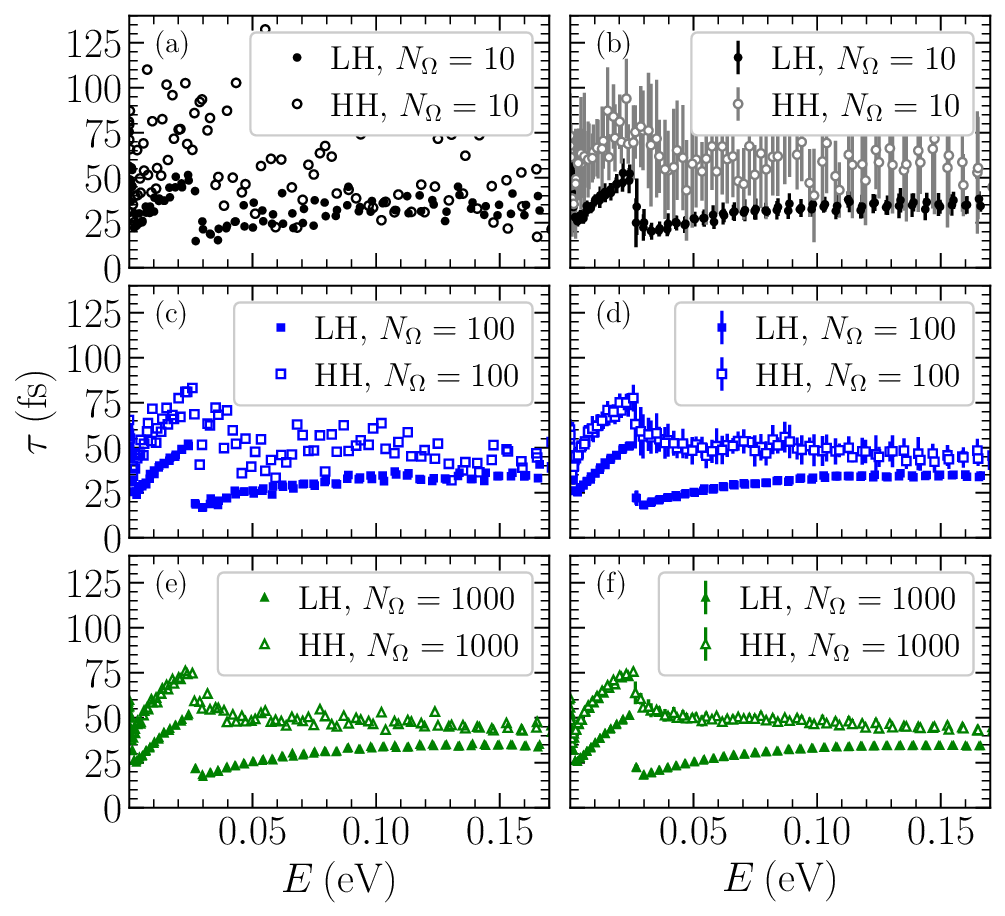}
\caption{Dependence of momentum relaxation time on the energy of a valence band \ac{HH} and \ac{LH} in ZnTe for three different values of the number of points $N_{\Omega}$ used in the calculation at $T=300\:\mathrm{K}$. The hole momentum is along the $\Gamma-L$ direction. The values shown in parts (a), (c) and (e) are the results for one realization.
The values shown in parts (b), (d) and (f) are the averages over ten different realizations, while the error bars are the standard deviation from these realizations. When error bars cannot be seen they are smaller than the symbol size. The results are presented for two nearly degenerate \ac{HH} and two nearly degenerate \ac{LH} bands.}\label{fig:a12}
\end{figure*}

\begin{figure*}[t]
\centering
\includegraphics[width=0.65\textwidth]{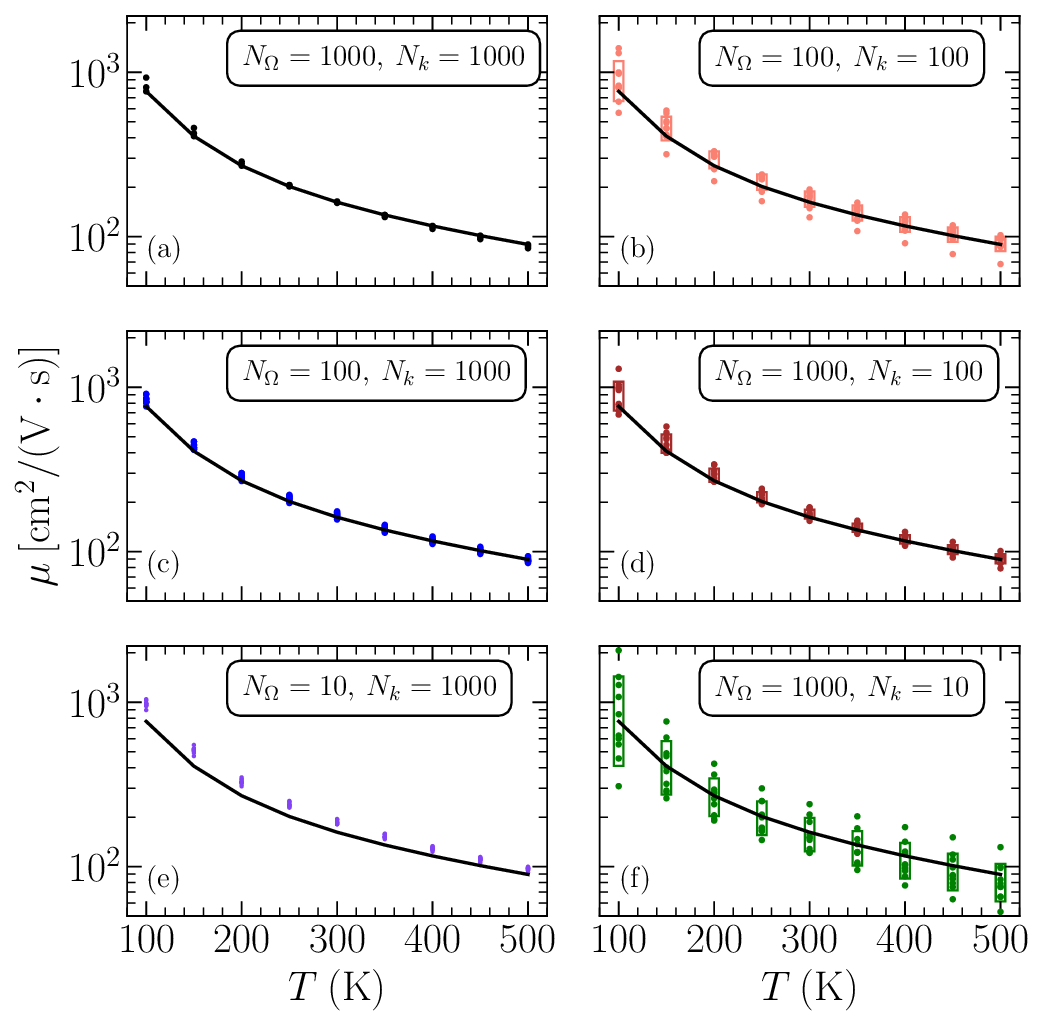}
\caption{Temperature dependence of hole mobility in ZnTe for different values of the number of $n\vb{k}$ pairs $N_k$ and the number $N_\Omega$. The dots present the results for ten (three when $N_k=1000$, $N_\Omega=1000$) different initial random number seeds. The rectangles denote the range $\qty(\mu-\sigma,\mu+\sigma)$ where $\mu$ is the average value and $\sigma$ is the standard deviation. Full line denotes the result for $N_k=1000$, $N_\Omega=1000$.}\label{fig:a13}
\end{figure*}

\end{onecolumn}

\end{document}